# Simulated assessment of light transport through ischaemic skin flaps

## Authors

Mark Main[1]

Richard J J Pilkington[2] BDS, FDSRCS (Eng), MBBS, MRCS (Eng)

Graham M Gibson[3] PhD, MSc

Akhil Kallepalli[3] PhD, MSc`

## Affiliations

1. James Watt School of Engineering, University of Glasgow, Glasgow G12 8QQ, United Kingdom
2. Oral and Maxillofacial Surgery, Cumberland Infirmary, Carlisle, CA2 7HY, United Kingdom
3. School of Physics and Astronomy, University of Glasgow, Glasgow G12 8QQ, United Kingdom

## Contact Details

Mark Main: Mark.Main@glasgow.ac.uk

Richard J J Pilkington: r.pilkington@nhs.net

Graham M Gibson: Graham.Gibson@glasgow.ac.uk

Akhil Kallepalli: Akhil.Kallepalli@glasgow.ac.uk

## Corresponding Author

Dr Akhil Kallepalli, School of Physics and Astronomy, University of Glasgow, Glasgow G12 8QQ, United Kingdom

Contact email: Akhil.Kallepalli@glasgow.ac.uk

# Simulated assessment of light transport through ischaemic skin flaps

## Abstract

Currently, free flaps and pedicled flaps are assessed for reperfusion in post-operative care using colour, capillary refill, temperature, texture and Doppler signal (if available). While these techniques are effective, they are prone to error due to their qualitative nature. In this research, we explore using different wavelengths of light to quantify the response of ischaemic tissue. The assessment provides us with indicators that are key to our goal of developing a point-of-care diagnostics device, capable of observing reduced perfusion quantitatively. We set up a detailed optical model of the layers of the skin. The layers of the model are given appropriate optical properties of the tissue, with due consideration of melanin and haemoglobin concentrations. We simulate 24 models of healthy, perfused tissue and perfusion-deprived tissue to assess the responses when illuminated with visible and near-infrared wavelengths of light. In addition to detailed fluence maps of photon propagation, we propose a simple mathematical model to assess the differential propagation of photons in tissue; the optical reperfusion factor (ORF). Our results show clear advantages of using light at longer wavelengths (red, near-infrared) and the inferences drawn from the simulations hold significant clinical relevance. The simulated scenarios and results consolidate the belief of a multi-wavelength, point-of-care diagnostics device and inform its design for quantifying blood flow in transplanted tissue. The modelling approach is applicable beyond the current research, wherein other medical conditions that can be mathematically represented in the skin can be investigated. Through these, additional inferences and approaches to other point-of-care devices can be realised.

## Introduction

Free-tissue transfer has become the method of choice for reconstructing defects produced by the ablative treatment of cancer and, less commonly, trauma in the head and neck region. This is a reliable method of reconstruction, resulting in improved postoperative function and aesthetic appearance[1]. The ideal flap monitoring system should be harmless to the patient and the flap, while also being of low cost, rapidly responsive, accurate, reliable, and applicable to all types of flaps. The monitor should be equipped with a simple display that would allow even the least experienced member of the team to easily recognise circulatory impairment. Such attempts for a device have been made in the past, yet there remains a necessity to achieve all the above requirements in one system[2–6]. Many techniques for postoperative monitoring of free-tissue transfers, both invasive and non-invasive, have been developed and evaluated in clinical settings as well as in experimental free flap models. Unfortunately, the level of evidence for efficacy of these techniques in clinical practice is often very low and cost/benefit analyses are rare. The only accepted practice in this field seems to be the almost universal use of clinical bedside monitoring of free-tissue transfers. This is supported by the overwhelming impression from the literature reviewed that clinical monitoring is the most practical and possibly reliable mode of monitoring[7].

In this article, we use Monte Carlo simulations for assessing visible and near-infrared wavelengths' interaction with healthy and ischaemic skin tissue and present the clinical relevance of using optical technologies for free flap monitoring. The results of these simulations will assist in inferring design strategies for a non-contact and safe point-of-care diagnostics device.

## Methods

When light interacts with biological tissue, a combination of reflection, refraction, scattering, and absorption occur (Figure 1). For safely assessing optical methods for diagnostics prior to clinical trials, Monte Carlo (MC) simulations are widely used for simulating the interactions between light and tissue.

For our assessment, we built a seven-layer tissue model (Figure 1) representing the epidermis, dermis, and subcutaneous fat layers[8]. To account for varying skin types, we quantified the melanin concentration in the epidermis broadly into three categories combining consecutive classifications in the Fitzpatrick scale[9]. The dermis layer is modelled in two scenarios of healthy and ischaemic tissues, as is the case for failing skin flaps. The common optical properties are detailed in Table S1.

The optical properties of skin layers in literature are critically analysed and chosen from multiple studies[11–17]. The absorption coefficient of the base layer (without absorbers), $\mu_a^{skin}$ (Eq. (1)) is dependent on the wavelength of the incident light[11]. Subsequently, the contribution of melanin is accounted for through equation (2). Combined, equation (3) gives the absorption coefficient of the epidermis layer. This is calculated for three scenarios of melanin volume fractions, i.e. ($f_{melanin}$ = 0.037, 0.134, 0.305). The absorption coefficient of the stratum corneum is calculated using the equation (4), that accounts for absorption due to water, baseline skin and wavelength. Further, adopted from previous studies, the scattering properties of the epidermal layers are assigned to the models and detailed in Table S2 and Table S3. The concentration of water is denoted by $C_{H_2O}$.

$$\mu_a^{skin} = 7.84 \times 10^7 \times \lambda^{-3.255} \ [\lambda: nm] \tag{1}$$

$$\mu_a^{melanin} = 6.6 \times 10^{10} \times \lambda^{-3.3} \ [\lambda: nm] \tag{2}$$

$$\mu_a^{epidermis} = f_{melanin} \mu_a^{melanin} + (1 - f_{melanin}) \mu_a^{skin} \tag{3}$$

$$\mu_a^{stratum\ corneum} = \left((0.1 - 0.3 \times 10^{-4}\lambda) + 0.125\mu_a^{skin}\right)\left(1 - C_{H_2O}\right) + C_{H_2O}\mu_a^{H_2O} \tag{4}$$

Mathematically, the absorption coefficient of the dermis layer is calculated as a total of its constituents. This includes the absorption due to water $\mu_a^{H_2O}$, oxygenated $\mu_a^{HbO_2}$ and deoxygenated $\mu_a^{Hb}$ blood, volume fraction of haemoglobin ($\gamma$) and the blood oxygen saturation ($S$). The absorption coefficient for the dermis layer, $\mu_a^{dermis}(\lambda)$, is defined as:

$$\mu_a^{dermis}(\lambda) = (1-S)\,\gamma\,C_{blood}\,\mu_a^{Hb}(\lambda) + S\,\gamma\,C_{blood}\,\mu_a^{HbO_2}(\lambda) + (1 - \gamma\,C_{blood})\,C_{H_2O}\,\mu_a^{H_2O}(\lambda) + (1 - \gamma\,C_{blood})(1 - C_{H_2O})\,\mu_a^0(\lambda) \quad (5)$$

The volume fractions of blood and water, $C_{blood}$ and $C_{H_2O}$, are adopted from Meglinski and Matcher (2002)[8]. The constituents of blood are quantified using $\gamma$; a factor determined by hematocrit ($Ht$), volume fraction of red blood count ($F_{RBC}$) and the volume fraction of haemoglobin in individual red blood cells ($F_{Hb}$). This is defined as $\gamma = Ht \times F_{RBC} \times F_{Hb}$. The value for $\gamma$ is kept consistent throughout all layer calculations with $Ht$ = 0.45, $F_{RBC}$ = 0.99 and $F_{Hb}$ = 0.25 as this represents typical values found in healthy tissue.

A failing skin flap experiences primarily a reduction in blood flow. To model this, the $C_{blood}$ values for layers are reduced by half to 0.02, 0.15, 0.02 and 0.05. Using equation (5), the optical properties of an ischaemic flap are calculated (Table S4, Table S5). Finally, the optical properties assigned to the subcutaneous fat layer are directly adopted from literature[8].

The seven-layer tissue models consider three scenarios of melanin distribution and two scenarios of blood flow. Each model is assessed by illuminating with a billion photons at four distinct wavelengths of 480 nm (blue), 520 nm (green), 650 nm (red) and 950 nm (near-infrared). Combined, our study presents a detailed assessment of photon propagation in a total of 24 models. Each model and the corresponding results are accessible online[20].

$$ORF_{1-2} = \left|\frac{\lambda_1 - \lambda_2}{\lambda_1 + \lambda_2}\right| \quad (6)$$

After the simulations are complete, we define a new mathematical factor to analyse the results. The optical reperfusion factor (ORF, Eq. (6)) will be used to quantify the differential interactions of light with healthy and ischaemic tissue. The experimental analogy of clinical relevance here is every flap site will include the flap itself, and the surrounding healthy tissue. The corresponding measurements at multiple wavelengths from different sites is possible for clinical use.

## Results and Discussions

By comparing each of the four wavelengths used in the simulations, substantial variation can be seen in tissue absorption and depth penetration (Figure S1). For wavelengths 480 nm and 520 nm, there is a high degree of absorption at beam incidence, which results in a lower degree of penetration to deeper layers. A few photons do propagate but these do not hold any significance in a clinical/experimental context. This is due to the high energy deposition in the upper layers which could cause photo-chemical and photo-biological damage under conditions of prolonged exposure or when using high power light sources.

When considering longer wavelengths such as red and near-infrared, photons propagate deeper into the tissue. Therefore, the dermis layer with varied blood perfusion characteristics can be investigated. The lateral absorption when using red light is reduced when compared to blue and green wavelengths. With sufficient penetration and low lateral absorption in both healthy and unhealthy tissue, this suggests red light at 650 nm would be an applicable wavelength to use when imaging a patch of tissue without causing any damage at skin-safe powers. Similarly, near-infrared light has both high penetration and lower lateral absorption, indicating it could be useful for imaging tissue patches (Figure S1).

Individually, each of the wavelengths provide inferences regarding photon propagation in different conditions of melanin distribution and/or blood perfusion. However, in a clinical context, a single wavelength system cannot achieve a quantitative differentiation of healthy and ischaemic tissue because the variations of absorption from the skin site can occur due to many reasons other than only blood flow changes. Therefore, to quantitatively assess the variation of blood flow in the flap, multiple wavelengths and/or different measurements must be taken. The measurements taken will be assessed by the proposed optical reperfusion factor (*ORF*), given in eq. (6). The normalising factor considers various combinations of measurements at different wavelengths.

When comparing the ORF maps (Figure S2), substantial variation can be seen further away from the point of incidence (a–f). These variations imply a high degree of absorption when considering these combinations, relatively, and therefore result in greater differences. For instance, deeper penetration of near-infrared wavelength compared to surface absorption of green photons results in a higher difference deeper in the tissue model. The variation in absorption between red and near-infrared is less significant at incidence, as indicated by a small region of low fluence.

Key differences can be seen in Figure S2 (g), (h). These compare combinations of healthy and ischaemic tissue models in near-infrared and red wavelengths. Promisingly, notable differences are evident in both cases. In Figure S2 (g), the variation in energy deposition is not significant beyond where the beam enters the tissue model, i.e., at the site of incidence. In Figure S2 (h), comparing healthy and ischaemic conditions under red light illumination, there is far greater variation in energy deposition. There are two distinct regions within this model which demonstrates the high absorption rate of red light by haemoglobin. The central column has little variation in energy deposition which is expected due to the smaller distance travelled, reducing the absorption, and scattering of the light. Further, melanin does not absorb red light as much as blue and green. Therefore, high depth penetration is seen in red wavelengths and beyond. The area to note is the region of higher difference surrounding this central column. As the distance increases from the entry point of the incident beam, the variation in energy deposition increases. This demonstrates the results of a difference in blood perfusion within tissue as the variance in absorption increases as a function of distance from the entry point of the incident beam.

To highlight the result of significance, we focus on the case of observing healthy and ischaemic blood perfusion under red illumination (Figure S2). A linear increase of variance is observed in the ORF map. A combination of red and near-infrared wavelengths illumination and measuring the absorption within healthy and ischaemic tissue is the key takeaway from the simulated, Monte Carlo assessments.

## Conclusions

The purpose of this research was to investigate whether optical methods could be used to determine varying blood volumes within a section of skin tissue. By varying the optical properties of tissue, taking account of a reduction blood volume, Monte Carlo simulations were used to analyse a seven-layer tissue model. The results of these simulations[20] detail both the interaction of light in this tissue as well as comparisons made between varying wavelengths and blood perfusion. Analysis of the data shows that across wavelengths and with single wavelengths, a variation can be found between healthy and ischaemic tissue. The variation found using the optical reperfusion factor is sufficient to allow for an experimental study to be conducted. The inferences from this study will be the input for developing a multi-wavelength system using red and near-infrared light to monitor the reperfusion. This can be done by collecting measurements from within and beyond the flap boundaries. The measurements collected from the surrounding tissue will act as a healthy reference to a potentially ischaemic free flap. This results in a quantitative analysis of blood perfusion within the transplanted tissue using a non-contact and safe point-of-care diagnostics device.

## Acknowledgements

We wish to acknowledge the support from the Research Fund awarded by the Endowments Sub Committee of the British Association of Oral and Maxillofacial Surgeons (2021), EPSRC Research Council funding [EP/T517896/1, EP/M01326X/1] and the Royal Society. We would also like to thank Dr Qianqian Fang, Shijie Yan and Dr Yaoshen Yuan (Computational Optics and Translational Imaging Lab, College of

## Data Availability

The MATLAB codes for each model and the full set of results are available online[20].

## Conflict of interest

We have no conflicts of interest to declare.

## Ethics statement/confirmation of patients' permission

Not applicable.

## Tables and Figures

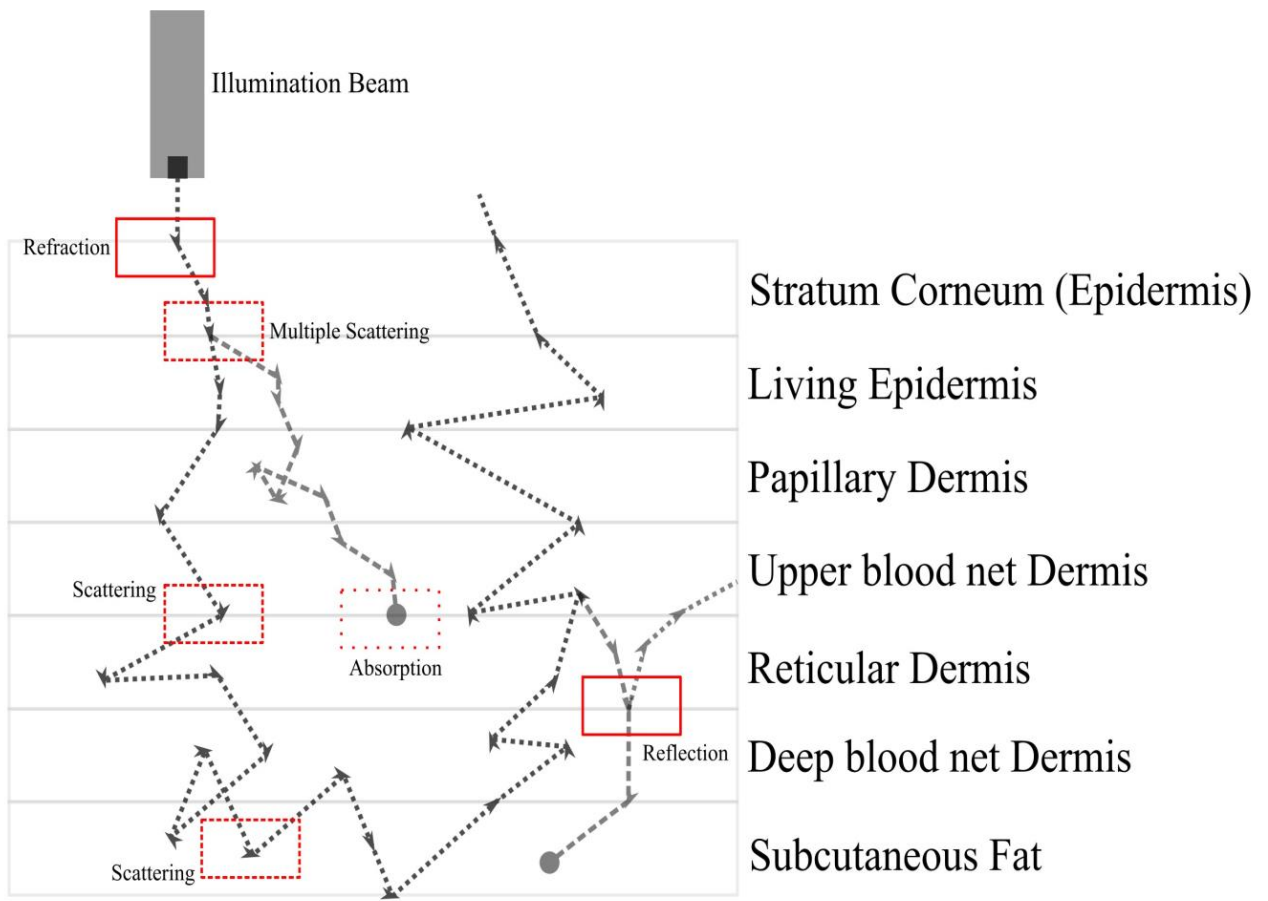

*Figure 1 – Light encounters multiple interactions when propagating through biological tissue. These include absorption, scattering and reflection at or within individual layers. This illustration shows an example of photon propagating through a multi-layer skin model. Each layer in the model is described by physical thickness (d), and optical properties, such as refractive index (η), absorption coefficient ($μ_a$), scattering coefficient ($μ_s$) and anisotropy (g). The model is adopted from Meglinski and Matcher (2002)[8].*

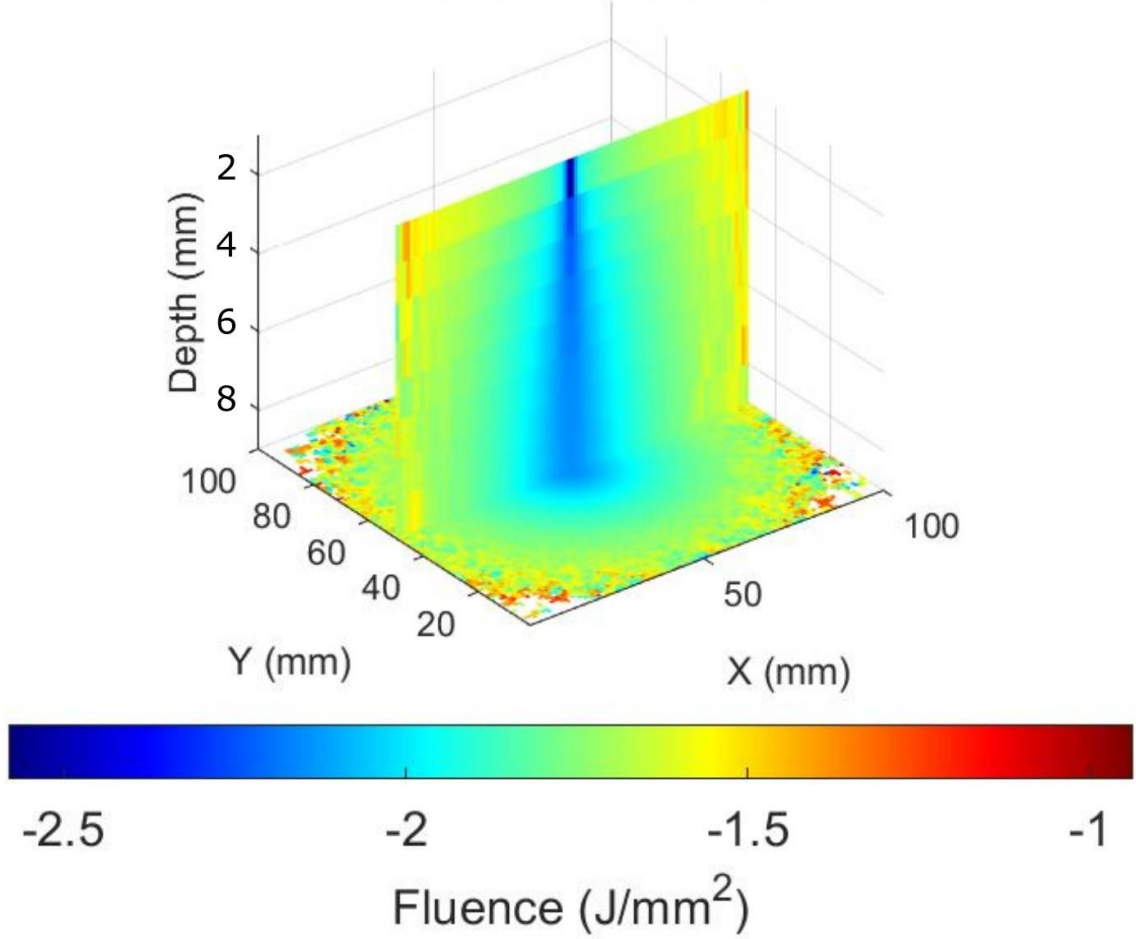

*Figure 2 – The optical reperfusion factor is calculated for red wavelength illumination of healthy and ischaemic tissue. Equation 6 was used to create the comparison and show the variation between the results. Healthy dermal perfusion is denoted with H, while ischaemic tissue models are denoted with U.*

## Supplementary Tables and Figures

*Table S1 - The tissue model is defined using the physical and optical properties of individual layers. The physical thickness (d), refractive index (η) and anisotropy (g) remain constant for all 24 models.*

| Layer | Thickness ($\mu$m) | Refractive index ($\eta$) | Anisotropy ($g$) |
|---|---|---|---|
| Stratum Corneum | 20 | 1.5 | 0.86 |
| Living Epidermis | 80 | 1.34 | 0.8 |
| Papillary Dermis | 150 | 1.4 | 0.9 |
| Upper blood net Dermis | 80 | 1.39 | 0.95 |
| Reticular Dermis | 1500 | 1.4 | 0.8 |
| Deep blood net Dermis | 100 | 1.38 | 0.95 |
| Subcutaneous fat | 6000 | 1.44 | 0.75 |

Table S2 – The absorption coefficients calculated from the relationships (eq. 1 – 4) for the epidermal layers are given below. Adopted from[8,11].

| Layer | Skin Types (I, II) | | | | Skin Types (III, IV) | | | | Skin Types (V, VI) | | | |
|---|---|---|---|---|---|---|---|---|---|---|---|---|
| | 480 nm | 520 nm | 650 nm | 950 nm | 480 nm | 520 nm | 650 nm | 950 nm | 480 nm | 520 nm | 650 nm | 950 nm |
| Stratum Corneum | 0.083 | 0.081 | 0.077 | 0.087 | 0.083 | 0.081 | 0.077 | 0.087 | 0.083 | 0.081 | 0.077 | 0.087 |
| Living Epidermis | 0.361 | 0.277 | 0.133 | 0.038 | 1.268 | 0.973 | 0.466 | 0.133 | 2.862 | 2.197 | 1.052 | 0.301 |

Table S3 – The scattering coefficients ($\mu_s$) adopted in this study for stratum corneum[8] and the living epidermis[11,17,18] are given below. The scattering coefficient for the living epidermis is calculated indirectly from the reduced scattering coefficient, $\mu_s'$, (given in the literature) using the anisotropy (g), as determined by Karsten and Smit (2012)[19].

| Layer | Skin Types (I, II) | | | | Skin Types (III, IV) | | | | Skin Types (V, VI) | | | |
|---|---|---|---|---|---|---|---|---|---|---|---|---|
| | 480 nm | 520 nm | 650 nm | 950 nm | 480 nm | 520 nm | 650 nm | 950 nm | 480 nm | 520 nm | 650 nm | 950 nm |
| Stratum Corneum | 45.889 | 41.444 | 28.389 | 12.889 | 45.889 | 41.444 | 28.389 | 12.889 | 45.889 | 41.444 | 28.389 | 12.889 |
| Living Epidermis | 45.889 | 41.444 | 28.389 | 12.889 | 45.889 | 41.444 | 28.389 | 12.889 | 45.889 | 41.444 | 28.389 | 12.889 |

*Table S4 – The absorption coefficient of the dermal layers is measured with contributions from absorption due to water, haemoglobin and deoxyhaemoglobin while accounting for the concentration of water, blood and its constituents. The properties of blood and haemoglobin is intrinsically included and modified, in the second dermis scenario, to represent ischaemic tissue.*

| Layer | Healthy | | | | Ischaemic | | | |
| --- | --- | --- | --- | --- | --- | --- | --- | --- |
| | 480 nm | 520 nm | 650 nm | 950 nm | 480 nm | 520 nm | 650 nm | 950 nm |
| Papillary Dermis | 0.059 | 0.071 | 0.009 | 0.196 | 0.034 | 0.039 | 0.006 | 0.195 |
| Upper blood net Dermis | 0.399 | 0.494 | 0.035 | 0.243 | 0.203 | 0.25 | 0.02 | 0.238 |
| Reticular Dermis | 0.057 | 0.069 | 0.008 | 0.273 | 0.031 | 0.036 | 0.006 | 0.272 |
| Deep blood net Dermis | 0.136 | 0.167 | 0.014 | 0.275 | 0.07 | 0.085 | 0.009 | 0.273 |

Table S5 – The scattering coefficients ($\mu_s$) adopted in this study for the dermal layer are calculated indirectly from the reduced scattering coefficient, $\mu_s'$, (given in the literature[13]) using the anisotropy (g), as determined by Karsten and Smit (2012)[19].

| Layer | 480 nm | 520 nm | 650 nm | 950 nm |
|---|---|---|---|---|
| Papillary Dermis | 53.8 | 44 | 26 | 15.9 |
| Upper blood net Dermis | 53.8 | 44 | 26 | 15.9 |
| Reticular Dermis | 53.8 | 44 | 26 | 15.9 |
| Deep blood net Dermis | 53.8 | 44 | 26 | 15.9 |

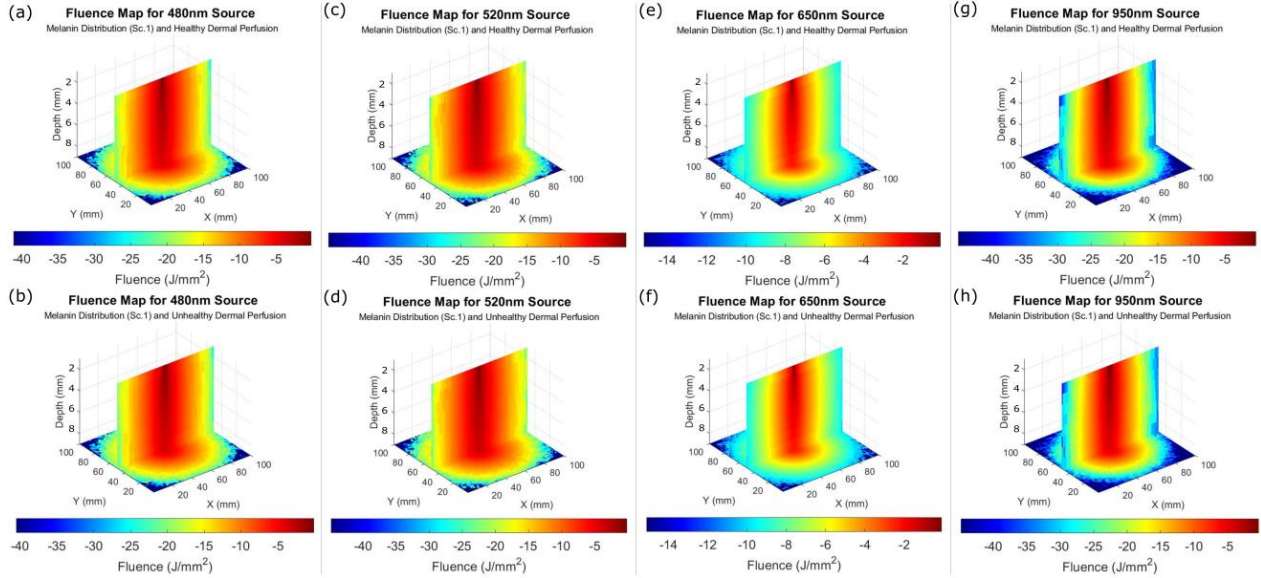

*Figure S1 – The results show different tissue model scenarios, simulated after interaction with 1 billion photons. The models have a constant melanin distribution (fmelanin = 0.037). The fluence maps show wavelengths and individual blood conditions attributed to the dermis layer. (a) 480 nm, healthy model. (b) 480 nm, ischaemic model. (c) 520 nm, healthy model. (d) 520 nm, ischaemic model, (e) 650 nm, healthy model, (f) 650nm, ischaemic model. (g) 950 nm, healthy model. (h) 950 nm, ischaemic model.*

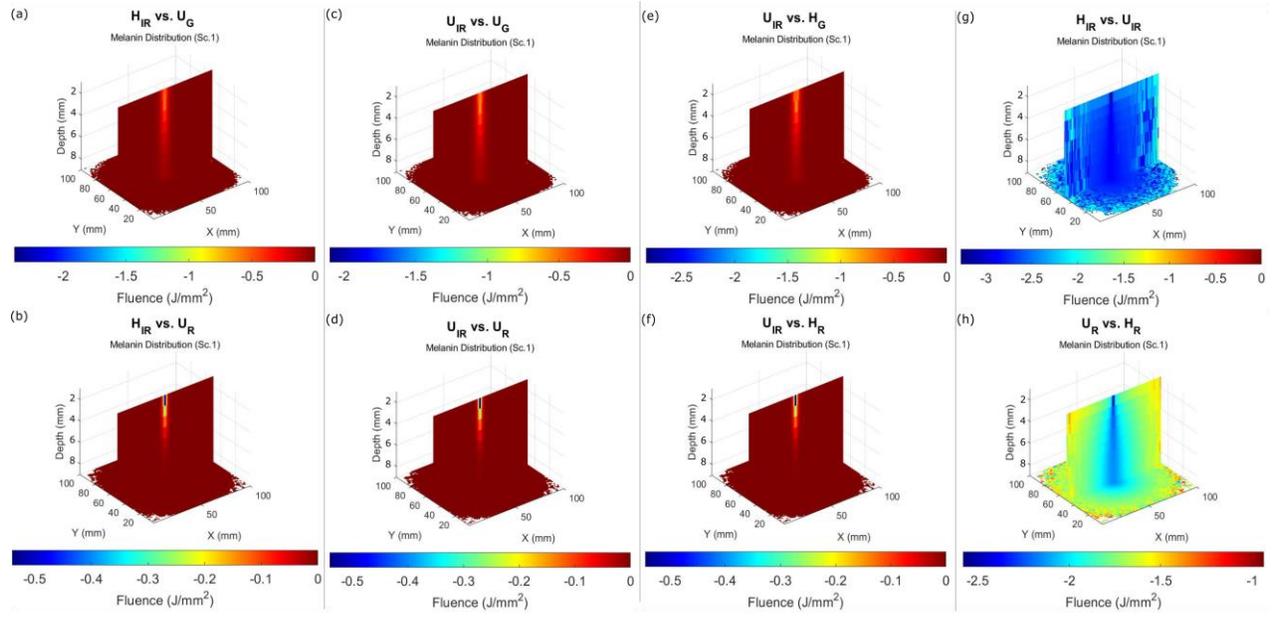

*Figure S2 – The optical reperfusion factor results are illustrated for tissue models with $f_{melanin}$ = 0.037. The results compare ORF maps for wavelength-tissue model combinations of (a) IR-healthy against green-ischaemic, (b) IR-healthy against red-ischaemic, (c) IR-ischaemic against green-ischaemic, (d) IR-ischaemic against red-ischaemic, (e) IR-ischaemic against green-healthy, (f) IR-ischaemic against red-healthy, (g) IR-healthy against IR-ischaemic and (h) red-ischaemic against red-healthy. To produce these ORF maps, equation 6 was used. Healthy dermal perfusion is denoted with H, while ischaemic tissue models are denoted with U.*